\documentclass[10pt]{iopart}
\usepackage{graphicx}
\expandafter\let\csname equation*\endcsname\relax
\expandafter\let\csname endequation*\endcsname\relax
\usepackage{amsmath} 
\usepackage{color,soul}
\usepackage{iopams}  
\begin{document}

\title[Effect of heteroepitaxial growth on LT-GaAs: ultrafast optical propertiess]{Effect of heteroepitaxial growth on LT-GaAs: ultrafast optical properties }

\author{Jessica Afalla$^{1}$, Elizabeth Ann Prieto$^{2,3}$, Horace Andrew Husay$^2$, Karl Cedric Gonzales$^{2,4}$, Gerald Catindig$^2$, Aizitiaili Abulikemu$^1$, Armando Somintac$^{2,3}$, Arnel Salvador$^{2,3}$, Elmer Estacio$^{2,3}$, Masahiko Tani$^5$ and Muneaki Hase$^1$}

\address{$^1$Faculty of Pure and Applied Sciences, University of Tsukuba, Tsukuba, Japan}

\address{$^2$ National Institute of Physics, University of the Philippines Diliman, Quezon City, Philippines}
\address{$^3$MSEP - College of Science, University of the Philippines Diliman, Quezon City, Philippines}
\address{$^4$Institute of Advanced Materials, Universitat Jaume I, Castelló, Spain}
\address{$^5$Research Center for Development of Far Infrared Region, University of Fukui, Fukui, Japan}
\ead{afalla.castillo.gn@u.tsukuba.ac.jp}
\vspace{10pt}
\begin{indented}
\item[]March 2021
\end{indented}

\begin{abstract}
Epitaxial low temperature grown GaAs (LT-GaAs) on silicon (LT-GaAs/Si) has the potential for terahertz (THz) photoconductive antenna applications. However, crystalline, optical and electrical properties of heteroepitaxial grown LT-GaAs/Si can be very different from those grown on semi-insulating GaAs substrates (“reference”). In this study, we investigate optical properties of an epitaxial grown LT-GaAs/Si sample, compared to a reference grown under the same substrate temperature, and with the same layer thickness. Anti-phase domains and some crystal misorientation are present in the LT-GaAs/Si. From coherent phonon spectroscopy, the intrinsic carrier densities are estimated to be ~$10^{15}$ cm$^{-3}$ for either sample. Strong plasmon damping is also observed. Carrier dynamics, measured by time-resolved THz spectroscopy at high excitation fluence, reveals markedly different responses between samples. Below saturation, both samples exhibit the desired fast response. Under optical fluences $\geq$ 54 $\mu$ J/cm$^2$, the reference LT-GaAs layer shows saturation of electron trapping states leading to non-exponential behavior, but the LT-GaAs/Si maintains a double exponential decay. The difference is attributed to the formation of As-As and Ga-Ga bonds during the heteroepitaxial growth of LT-GaAs/Si, effectively leading to a much lower density of As-related electron traps.
\end{abstract}

%
\vspace{2pc}
\submitto{\JPCM}
%
%
\ioptwocol

\section{Introduction}

The integration of semiconductor layers on silicon substrates is an ongoing pursuit for crystal growers \cite{Petrushkov2020},\cite{Barrett2019},\cite{Bugomilowicz2016}, \cite{Wirths2018}. For GaAs on silicon (GaAs/Si), the difficulty in heteroepitaxial growth is attributed to the lattice mismatch, the growth of a polar material on a non-polar substrate, and the difference in thermal expansion coefficients [5]. When semiconductor devices are successfully grown on or transferred to silicon substrates, doing away with the more expensive substrates, such as GaAs, ideally saves device space (“monolithic”) and reduces production cost \cite{Barrett2019},\cite{Bolkhovityanov2008}. In the past, low temperature growth was explored to achieve high quality GaAs \cite{Gonzalez1992}. In more recent context, work has shown that epitaxial grown low temperature GaAs on silicon (LT-GaAs/Si) is a viable material for terahertz (THz) emission  \cite{CGonzales2021} and detection \cite{Afalla2020}. The device performance of the LT-GaAs/Si photoconductive antennas (PCAs) have been evaluated based on the samples’ crystalline and optoelectronic qualities \cite{Afalla2020}. A particular advantage of using silicon substrates for THz applications is the significant reduction in THz absorption by the substrate \cite{Kamo2014},\cite{Klier2015},\cite{Kasai2009}. The bandwidth of LT-GaAs PCAs is limited by the phonon modes of GaAs between 8-9 THz. 

For conventional LT-GaAs (on GaAs substrates), crystal properties, carrier dynamics and photoconductive properties that depend on growth conditions are well documented \cite{Kamo2014},\cite{Youn2008},\cite{Prahbu1997}. In addition to low temperature growth, an As-rich environment is favorable for the formation of As interstitials and As antisites. These point defects and related As precipitates (or clusters) create deep-level electronic traps in the LT-GaAs bandgap. The process of efficient carrier trapping into such As-related defects is responsible for the ultrafast response time of LT-GaAs PCAs \cite{Kamo2014}, \cite{Youn2008},\cite{Prahbu1997}. Although the introduction of defects in LT-GaAs is intentional, a certain level of crystallinity is required for efficient, high mobility photoconductive devices. The inequivalence of homo- and heteroepitaxial growth leads to LT-GaAs layers that have vastly different properties – even if they were grown under similar conditions (such as the same substrate temperature). 

In this work, we investigated the different optical and crystalline properties that result from heteroepitaxial growth of LT-GaAs on silicon , in contrast to a reference LT-GaAs layer grown on a GaAs substrate. Under high photoexcitation density, the samples exhibit markedly different carrier dynamics, which we attribute to the different defect profiles resulting from how the excess As environment was incorporated in the layers during growth. Carrier densities and phonon characteristics were obtained using coherent phonon spectroscopy, which has only been previously utilized for LT-GaAs grown on GaAs substrates \cite{Dekorsy1993}. Scanning electron microscopy and Raman microspectroscopy provided crystallographic information that supplemented the results of the optical measurements.

\section{Methods}

\subsection{Sample growth} 

\begin{figure}
	\centering
	\includegraphics[width=60 mm]{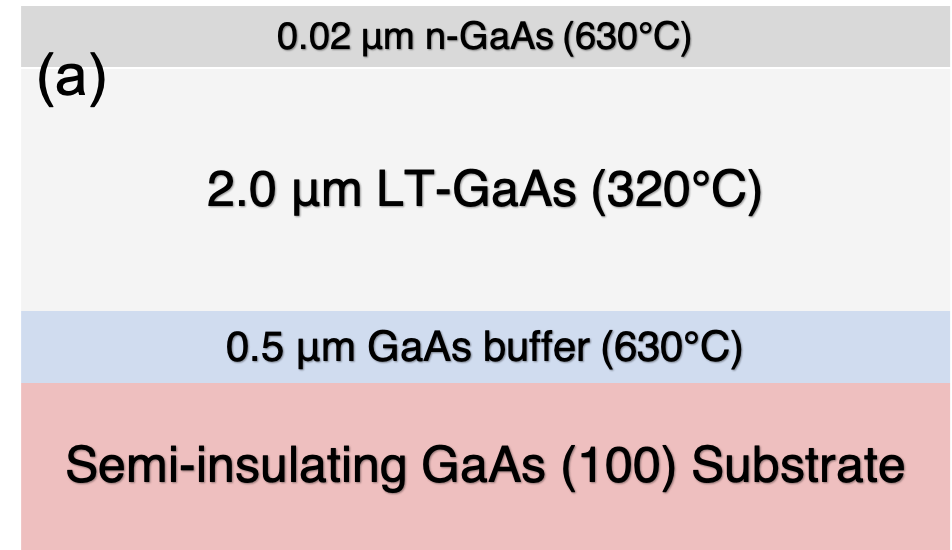}
	
	\vspace{5mm}
	
	\includegraphics[width=60 mm]{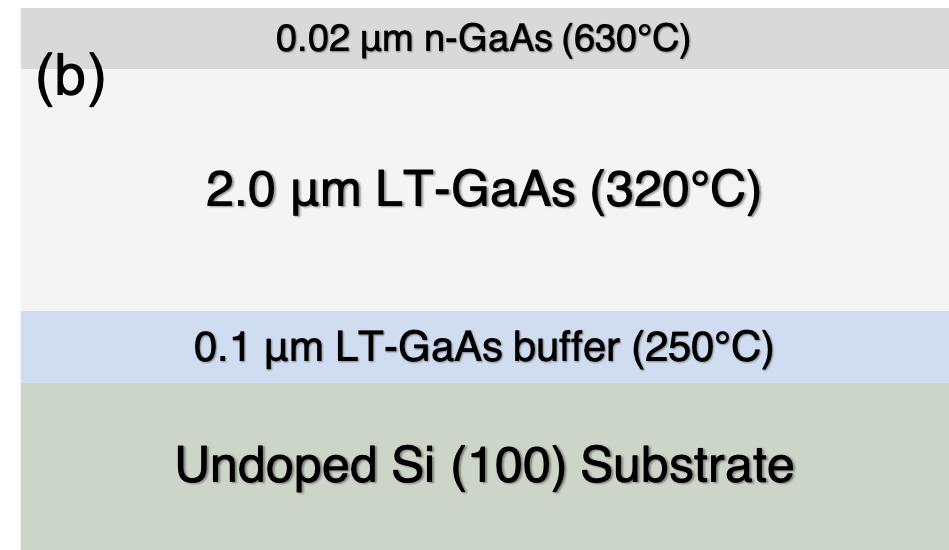}
	\caption{Layer schematic of LT-GaAs layers grown at 320ºC on different substrates, (a) SI-GaAs (100) and (b) Si (100). The bulk LT-GaAs and n-GaAs capping layers were grown similarly.}
	\label{fig:fig1}
\end{figure}

The samples used in this study were two LT-GaAs layers grown using a 32P Riber molecular beam epitaxy (MBE) facility under the same substrate temperature, 320ºC. The layer schematics are shown in Figure \ref{fig:fig1}. The MBE facility is equipped with solid Ga and As$_{4}$ sources. The samples were epitaxial grown on either an n-type undoped Si substrate, or a semi-insulating GaAs (SI-GaAs) substrate under similar As overpressure; both substrates were (100)-oriented. For the LT-GaAs/Si sample, the n-type undoped Si (100) substrate first underwent ionic cleaning and oxide stripping. And prior to any deposition, passivation layer removal was done in situ (700ºC for 10 mins). A thin LT-GaAs buffer was deposited, and subsequently, the growth of the 2.0 $\mu$m LT-GaAs layer was done at a substrate temperature of 320ºC. The growth was terminated by a 0.02 $\mu$m n-GaAs capping layer for future Ohmic contacts metallization. As these layers were intended for photoconductive antenna application, the n-GaAs cap also allows for more effective current conduction between fabricated metal contacts and the highly resistive LT-GaAs layer. In situ annealing for 10 mins at 600$^\circ$C was conducted each time an LT-GaAs layer is completed.

For the LT-GaAs/SI-GaAs sample (“reference”), a 0.5 $\mu$m GaAs buffer layer was initially deposited at 630ºC. Subsequently, a similar 2.0 $\mu$m-thick LT-GaAs layer was grown at 320ºC and was likewise terminated with a 0.02 $\mu$m n-GaAs cap. The reference sample was annealed in situ at 600$^\circ$C for 10 mins.

\subsection{Coherent phonon spectroscopy}
Coherent phonon spectroscopy was performed in order to estimate the samples’ carrier densities and to obtain phonon characteristics. Measurements were conducted using an optical pump-probe set-up, with a Ti:Sapphire oscillator providing 30 fs pulses at 830 nm wavelength and a repetition rate of 80 MHz. The cross-polarized pump and probe beams were focused on the sample surface with a spot size of ~8 $\mu$m at angles 5º (pump) and 15º (probe) with respect to the sample normal.  As the pump-probe delay was varied using a fast-scanning shaker delay line, the time-resolved reflectivity change, $\Delta$R/R, was measured (pump fluence ~850 $\mu$J/cm$^2$). As these samples were undoped, non-electro-optic detection was chosen to eliminate the large LO-phonon response. This enabled better access to the GaAs TO and/or longitudinal optical plasmon coupled (LOPC) for investigation. 

\subsection{Time-resolved THz spectroscopy}
The carrier dynamics as a function of photoexcitation density were investigated using time-resolved THz transmission, -$\Delta T/T$. It was performed using an optical pump-terahertz probe system. A Ti:Sapphire femtosecond laser amplifier system operating at a repetition rate of 1 kHz, emits 40 fs pulses with a wavelength of 800 nm. The THz radiation was generated from a two-color plasma, created by passing the laser through a $\beta$-BBO crystal. The THz probe was focused by a pair of parabolic mirrors and was incident normal to the sample surface. The transmitted THz wave was detected using a ZnTe crystal used in an electro-optic sampling scheme. The optical pump was also incident normal to the sample surface, and the pump-probe delay was varied using a controlled delay stage. The pump fluence was varied between $20$ and $120 \mu$J/cm$^2$, equivalent to $5\times10^{18}$ photons/cm$^3$ and $3.2 \times10^{19}$ photons/cm$^3$, respectively. 

\subsection{Raman micro-spectroscopy}
Raman micro-spectroscopy in (100) backscattering geometry was performed using a 532 nm continuous wave fiber laser excitation source. The laser was focused using a 50$\times$ objective lens on a spot close to the sample surface. 

\section{Results and Discussion}
\subsection{Surface quality}
\begin{figure}
	\centering
	\includegraphics[width=\linewidth]{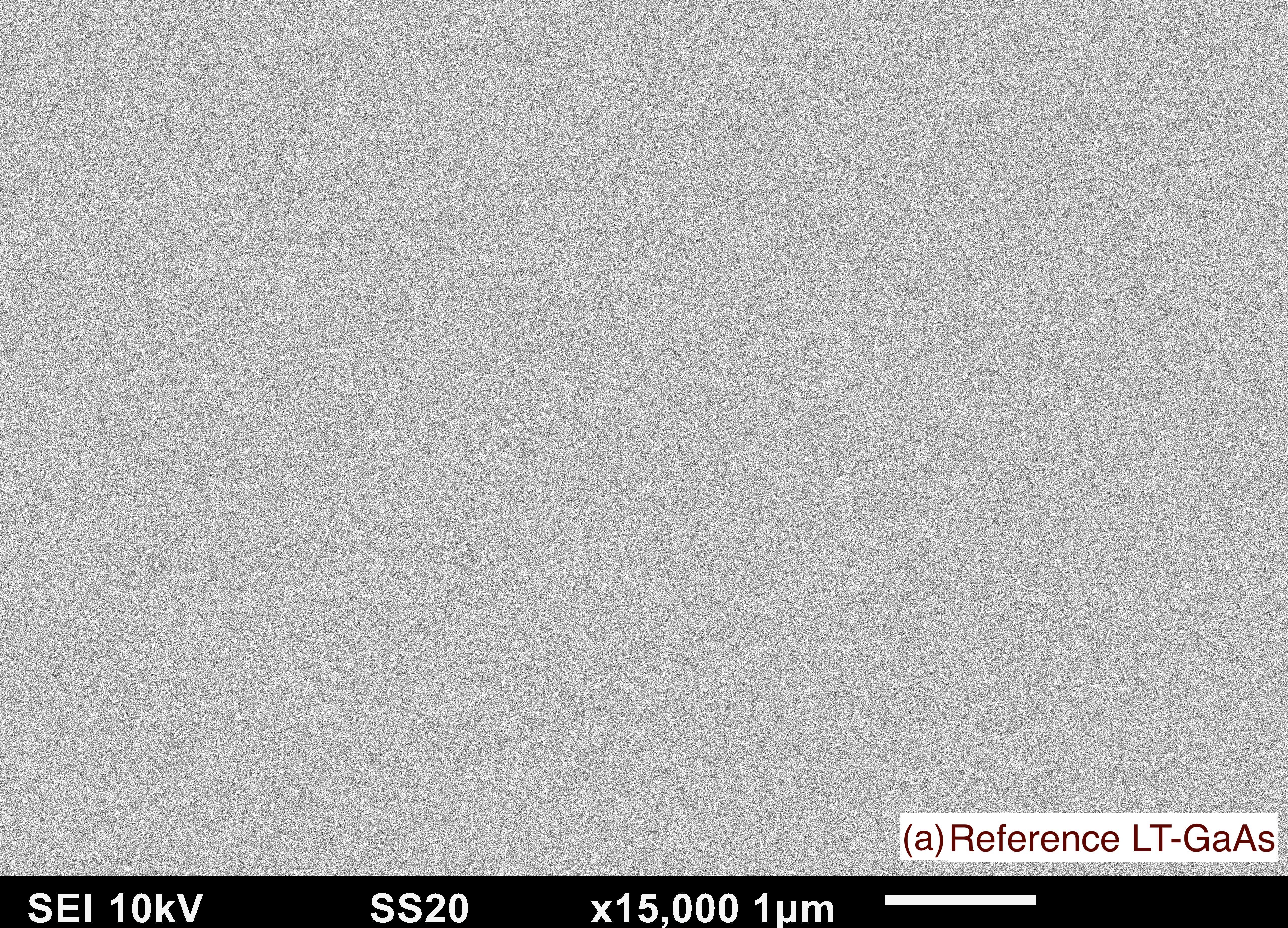}
	\includegraphics[width=\linewidth]{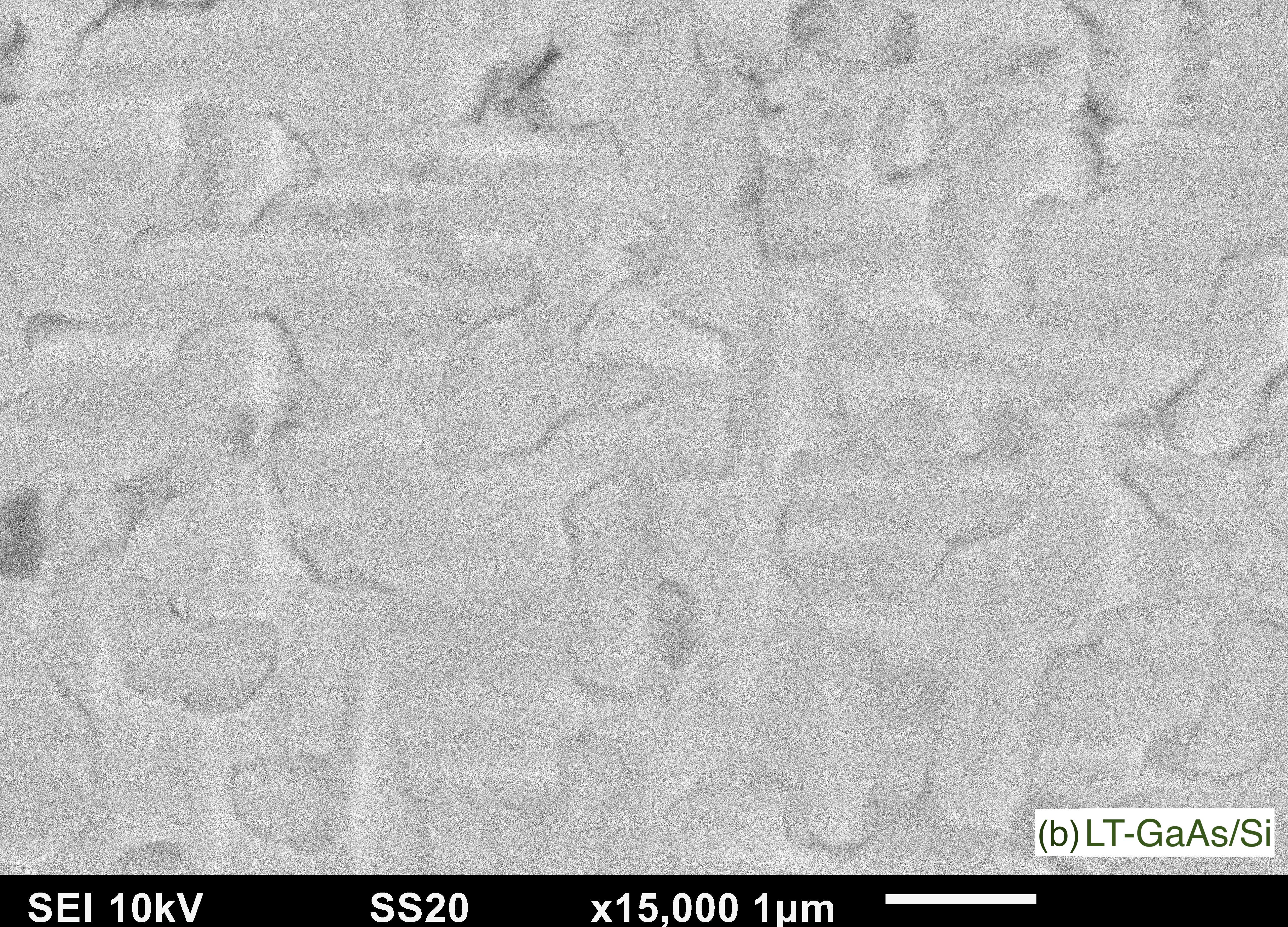}
	\caption{SEM images of the sample surfaces at 15,000x magnification, where the (a) reference LT-GaAs remains featureless. At high magnification, anti-phase domains appear as mosaic-like features on the surface of the (b)LT-GaAs/Si sample.}
	\label{fig:fig2}
\end{figure}
On visual inspection, the reference sample has a smooth, mirror-like surface. The surface of the LT-GaAs/Si sample has no cracks, holes or visible defects. Scanning electron micrographs (Figure \ref{fig:fig2}) at $15,000\times$ magnification reveal mosaic-like features on the sample surface. Similar features on surface images of GaAs grown on silicon substrates taken using SEM \cite{Georgakilas1992},\cite{Georgakilas1993} and AFM\cite{Alcotte2016} have been identified anti-phase domains. These mosaic-like features, (described as "facets" \cite{Georgakilas1992}) indicate anti-phase boundaries (APBs) in the bulk\cite{Georgakilas1992},\cite{Georgakilas1993},\cite{Alcotte2016}. Using nominal Si (100) substrates often lead to the creation of APBs during the growth process \cite{Barrett2019}, \cite{Koch1987}, which in turn yields these anti-phase domains on the surface. Under the same magnification, the surface of the reference sample remains featureless.

\subsection{Carrier density, phonon characteristics}
\begin{figure}
	\centering
	\includegraphics[width=\linewidth]{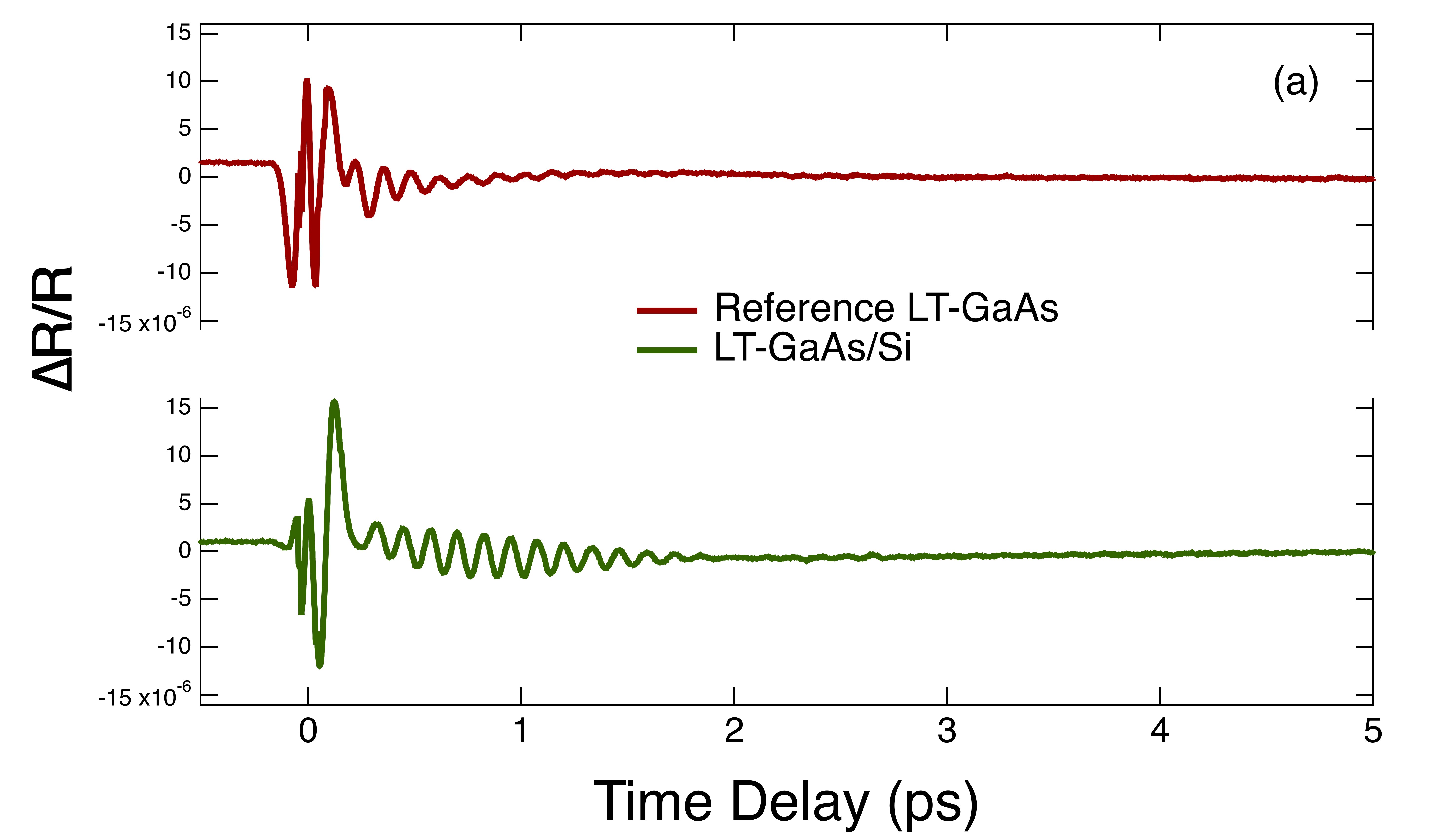}
	\includegraphics[width=\linewidth]{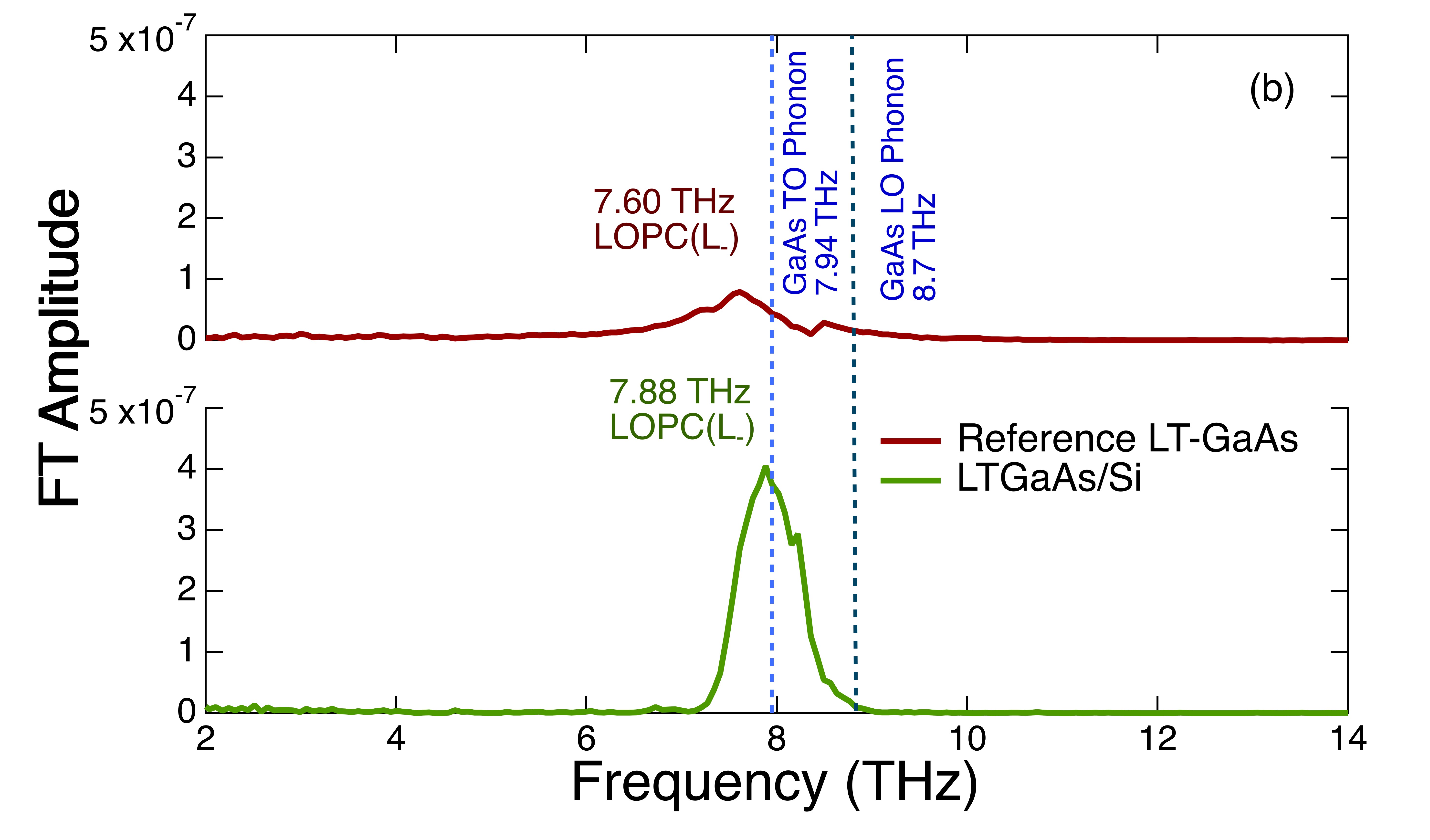}
	\caption{(a) Time resolved reflectivity change showing coherent phonon oscillations and (b) their corresponding Fourier-transformed amplitude spectra. The GaAs LO and TO phonon frequencies are indicated by the vertical dashed lines. }
	\label{fig:fig3}
\end{figure}

The $\Delta$R/R scans, plotted in Figure \ref{fig:fig3}a, show coherent phonon oscillations propagating within a few picoseconds for either sample. Figure \ref{fig:fig3}b shows the corresponding amplitude spectra. Under the fluence used, and a refractive index of $n=3.6$ for GaAs, the photoexcitation density was ~$1.06 \times 10^{18}$ cm$^{-3}$. The frequencies of the coherent phonon oscillations were found to be 7.88 THz for LT-GaAs/Si, and 7.60 THz for the reference, which are attributed to each sample’s L$_{-}$ branch of the LOPC modes \cite{Mooradian1967}. The former exhibits almost quadruple in amplitude. The absence of the L+ mode indicates that the samples’ intrinsic carrier densities are below $10^{16}$ cm$^{-3}$ \cite{Hase1998}.

\begin{figure}
	\centering
	\includegraphics[width=\linewidth]{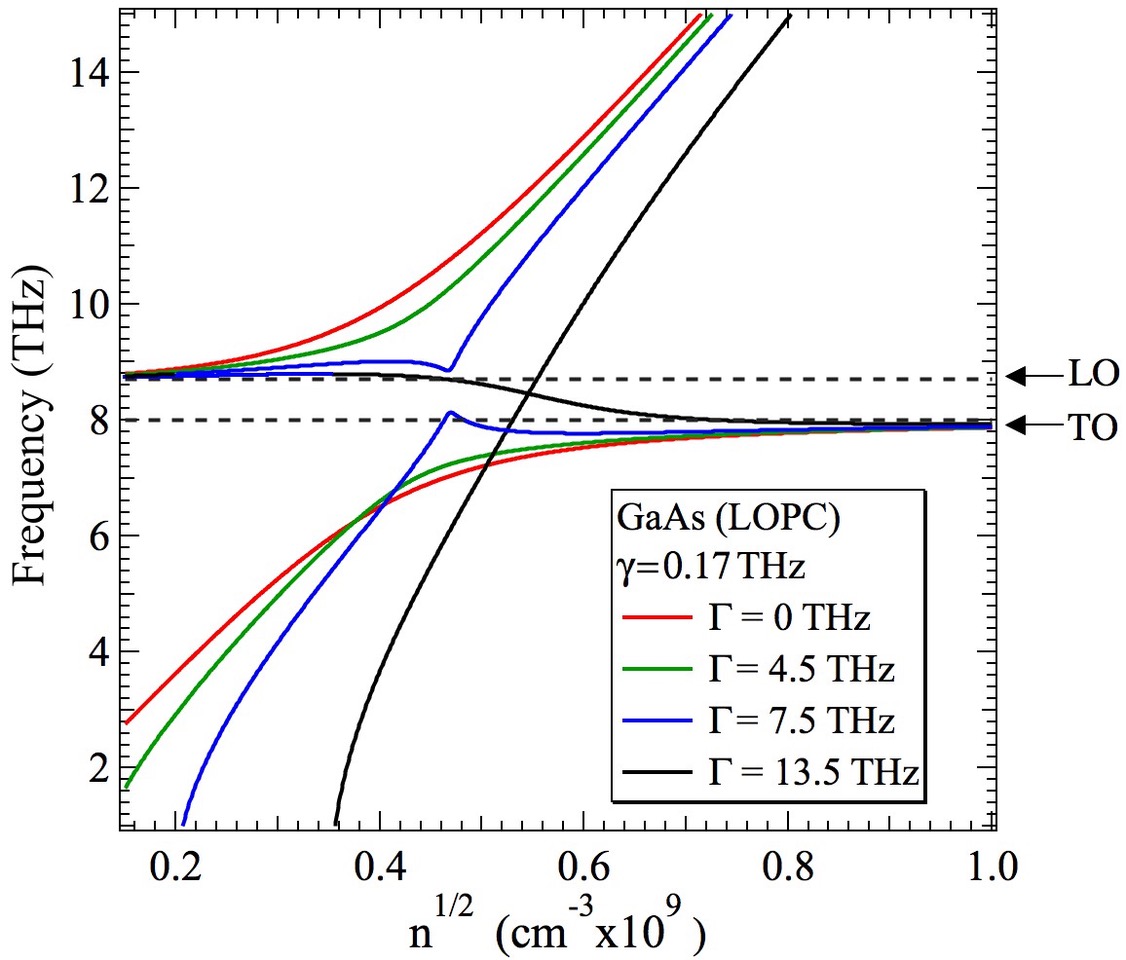}
	\caption{(Simulation of GaAs LO phonon-plasmon coupled (LOPC) modes with different plasmon damping constants, $\Gamma$. The LO phonon damping was fixed at $\gamma$ = 0.17 THz. The dashed lines show the GaAs LO and TO phonon frequencies.}
	\label{fig:fig4}
\end{figure}

Figure \ref{fig:fig4} shows a simulation of the GaAs LOPC L$_{-}$ and L$_{+}$ branches \cite{Fukusawa1994},\cite{Dekorsy2000} with varying plasmon damping constant, $\gamma$, under an assumed LO phonon damping constant of $\Gamma$ = 0.17 THz.  From the simulation curves, the location of the L$_{-}$ modes gives total carrier densities (intrinsic and photoexcited) of $1.0 \times 10^{15}$ cm$^{-3}$ for the LT-GaAs/Si and $0.8 \times 10^{18}$ cm$^{-3}$ for the reference. The total carrier density was therefore found to be slightly higher in the LT-GaAs/Si. Previously grown LT-GaAs samples from the same MBE facility have been measured to have intrinsic carrier densities of 10$^{15}$ cm$^{-3}$ when grown at 270ºC and annealed in situ at 600ºC \cite{Balgos2019}; or calculated to have photocarrier densities of up to $10^{17}$ cm$^{-3}$ when grown at 400ºC and annealed in situ at 600ºC \cite{Prieto2014}. Thus, the intrinsic carrier densities in the order of $10^{15}$ cm$^{-3}$ for these samples appear reasonable. This intrinsic carrier density is sufficient to observe the coherent phonon oscillations under high incident fluence, but still too low to be able to observe the L$_{+}$ of the LOPC \cite{Hase1998}. Furthermore, spectral features between the LO and TO phonon modes may be observed when the plasmon damping constant is high, as shown by $\Gamma$= 7.5 and 13.5 THz simulation curves in Figure \ref{fig:fig4}. For either amplitude spectra, as shown in Figure \ref{fig:fig3}b, a small kink-like feature is present, located at ~8.22 THz for LT-GaAs/Si and ~8.3 THz for the reference. 

\subsection{Carrier dynamics: presence and absence of trap states}
\begin{figure}
	\centering
	\includegraphics[width=\linewidth]{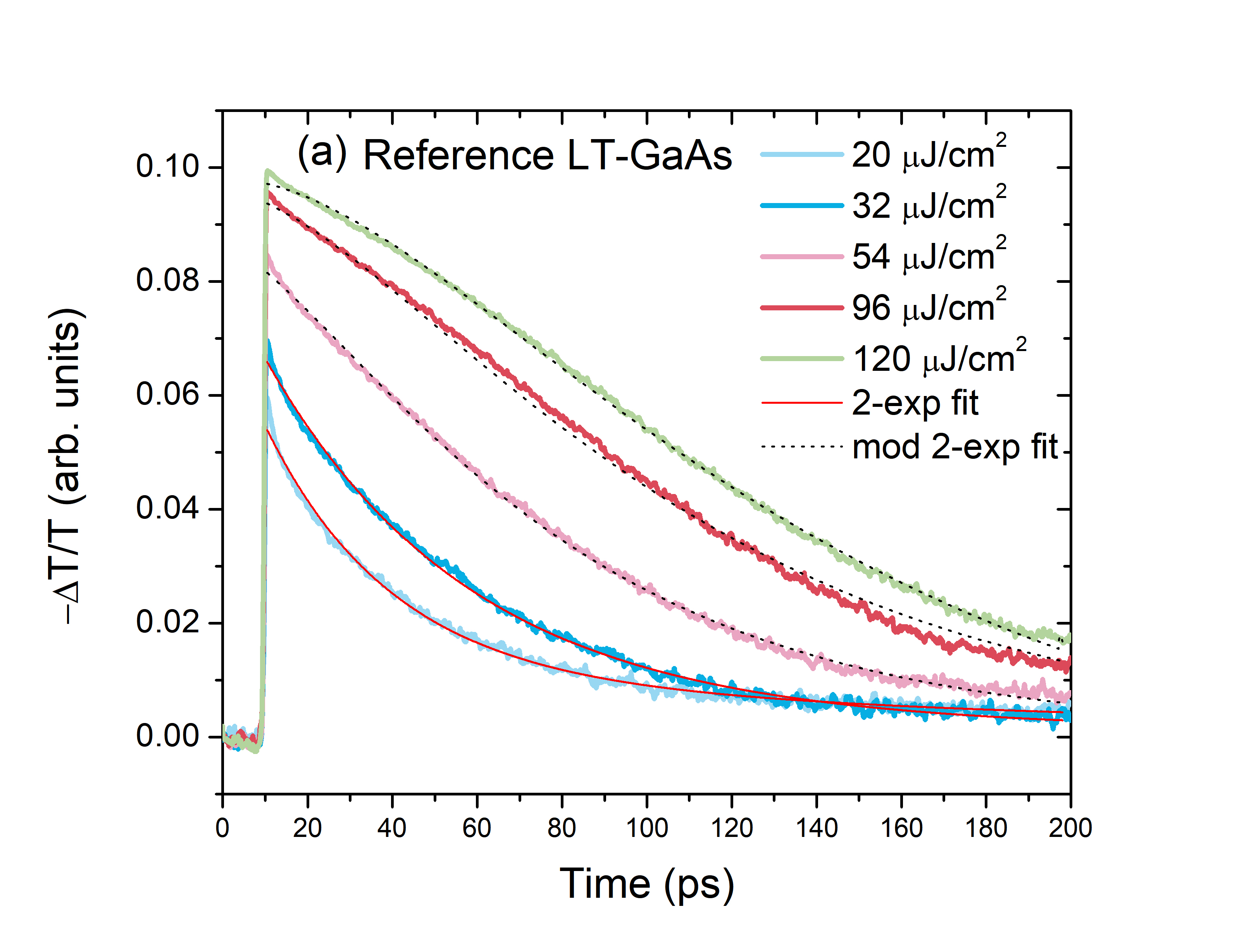}
	\includegraphics[width=\linewidth]{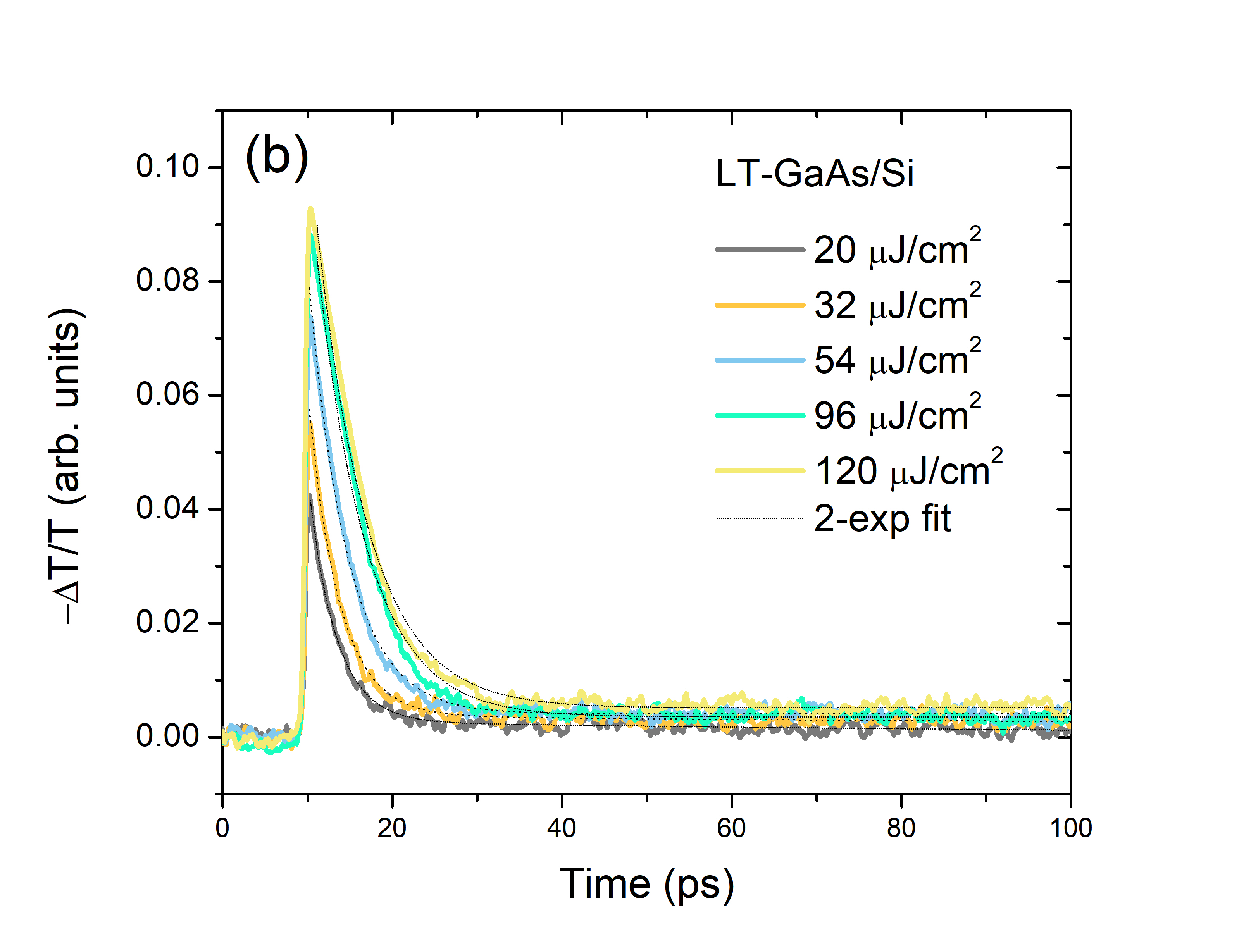}
	\caption{Time-resolved THz transmission curves of (a) the reference and (b) the LT-GaAs-on-Si layers under high excitation densities. As the pump fluence is increased, the behavior of (a) the reference -$\Delta$T/T deviates from a double-exponential decay. This is attributed to the filling, and possible saturation of defect states. Fitting of the decay curves are overlaid on the experimental data.}
	\label{fig:fig5}
\end{figure}

Figure \ref{fig:fig5} shows the time-evolution of the THz transmission, $-\Delta T/T$. As the pump arrives at the surface of either sample (set to time = 10 ps in this experiment), photoexcited carriers absorb the THz radiation, and THz transmission decreases within 1 ps. The maximum value for this change in transmission, $\left|\frac{\Delta T_{max}}{T}\right|$  increases linearly with fluence, and reaches a saturation. Through the use of a saturation equation in the form of 
\begin{eqnarray}
\left|\frac{\Delta T_{max}}{T}\right|=A\left(1+\frac{F_{sat}}{F}\right)\label{eq1}
\end{eqnarray}

where F is the fluence, A is a scaling factor, the fluence at which the transmission change saturates ($F_{sat}$) can be calculated. The saturation fluence $F_{sat}$ was calculated to be 19.6 $\mu$J/cm$^3$ for the reference and 37 $\mu$J/cm$^3$ for the LT-GaAs/Si. In this experiment, the fluence values used were $\geq 2$ orders of magnitude above the intrinsic carrier density.

After the initial decrease, $-\Delta T/T$, recovers (Figure \ref{fig:fig5}). For the reference at low fluence (below 54 $\mu J/cm^3 $) and for the LT-GaAs/Si at all fluence values, the carrier dynamics can be described as a double exponential decay–an initial fast process followed by a slow decay,

\begin{eqnarray}
N(t)=N_0\left(\gamma_\text{fast}e^{-\frac{t}{\tau_\text{fast}}}+\gamma_r e^{-\frac{t}{\tau_r}}\right) \label{eq2}
\end{eqnarray}
where $N(t)$ is the photoexcited electron population with initial value of $N_0$; $\gamma_\text{fast}$ and $\gamma_r$ are the fraction of the population participating in the fast initial process, and in recombination, respectively. The corresponding time constants are $\tau_\text{fast}$ and $\tau_r$. The second process, attributed to carrier recombination, manifests as a “pedestal” \cite{Segschneider2002} – wherein the transmission does not fully recover within the measurement window, and almost appears constant due to its very long decay time. Table 1 shows the decay times obtained using a double exponential decay fitting. The relatively long time constants for the reference sample is due to the growth temperature being above the usual 200-240ºC for commercial LT-GaAs antennas [7], as well as the high incident fluence \cite{Segschneider2002}. As the photoexcitation density increases, the decay time increases.

\begin{table}
\caption{\label{Table 1}Fast process time constants obtained from photocarrier decay fitting}
\footnotesize
\begin{tabular}{@{}lll}
\br
Fluence & LT-GaAs/Si & Reference\\
$\mu J/cm^3$&$\tau_\text{fast}$&$\tau_\text{trap}$\\
\mr
$20 \mu J/cm^3$&3.25 ps$^{\star}$&29.47 ps$^{\dagger}$\\
$32 \mu J/cm^3$&3.61 ps$^{\star}$&48.47 ps$^{\dagger}$\\
$54 \mu J/cm^3$&4.52 ps$^{\star}$&50.32 ps$^{\ast}$\\
$96 \mu J/cm^3$&5.76 ps$^{\star}$&69.50 ps$^{\ast}$ \\
$120 \mu J/cm^3$&6.00 ps$^{\star}$&83.04 ps$^{\ast}$\\
\br
\end{tabular}\\
$^{\star,\dagger,\ast}$ obtained using Equation 2,3,4 respectively

\end{table}
\normalsize

For conventional LT-GaAs, the initial fast process is attributed to carrier trapping in electronic defect states, such that,
\begin{eqnarray}
\gamma_{\text{fast}}e^{-\frac{t}{\tau_{\text{fast}}}}\rightarrow \gamma_{\text{trap}}e^{-\frac{t}{\tau_{\text{trap}}}} \label{eq3}
\end{eqnarray}
The defect density increases with decreasing growth temperature and/or annealing \cite{Segschneider2002},\cite{Nemec2001}, varying anywhere from 10$^{18}$ – 10$^{20}$ defects/cm$^3$. At 54 $\mu$J/cm$^3$ ($1.48 \times 10^{18}$ cm$^{-3}$) and higher, the shape of the reference $-\Delta T/T$ decay deviates from this double exponential decay, indicating that the electron traps began saturating. The group of G. Segschneider, et al have reported differential transmission curves behaving very similarly under high excitation conditions \cite{Segschneider2002}. The shapes of the curves were attributed to the saturation of defect states, which initially causes a slowing down of the trapping process. It was speculated that a trap-emptying process occurs simultaneously, with time constants that were characteristic of each sample. This interplay produces the “hump” seen in time evolution of -$\Delta T/T$. Using a decay rate equation, similar to Sosnowski, et al\cite{Sosnowski1997},

\begin{eqnarray}
N(t)=N_0(\gamma_\text{trap} e^{{-\frac{t}{\tau_\text{trap}}}}(1-\gamma_\text{empty} e^{{-\frac{t}{\tau_\text{empty}}}})+\gamma_re^{-\frac{t}{\tau_r}})\label{eq4}
\end{eqnarray}

the high excitation curves were fitted taking into consideration the effect of trap occupation through a generalized "emptying" term (see Supplementary Material for details). The observation of trap state saturation with increasing optical excitation in this study underscores the successful formation of As-related electron traps in the reference LT-GaAs layer. The results of -$\Delta T/T$ shows that the trap density is in the order of 10$^{18}$ cm$^{-3}$ for the reference.

For the LT-GaAs/Si (Figure \ref{fig:fig5}b), there is no trap state-saturation behavior observed. Regardless of incident fluence, the -$\Delta T/T$ can be described by a double exponential decay. Relative to the reference, the pedestal also appears lower (only $\leq$6$\%$ of the photocarrier population). Although the excitation density was quite high, the initial fast process had a time constant ($\tau_{\text{fast}}$) that remained within the same order of magnitude. At 20 $\mu$J/cm$^2$, $\tau_{\text{fast}}=3.25$ ps, and only increases to 6 ps when the fluence is at 120 $\mu$J/cm$^{2}$ (Table 1). In the presence of point defects such as $As_{Ga}$, or ion-implanted defects, as well as the As clusters that form upon annealing, increasing the photoexcitation density leads to the increase of trapping time, and a difference in amplitude ratio between the two exponential decay processes – trapping and carrier recombination, changes. The former is caused by the filling of trap states, and the latter is due to more carriers diffusing and recombining in the bulk effects which can be observed even at low photoexcitation densities \cite{Nemec2001}, \cite{Savard2010}, \cite{Sosnowski1997}. The absence of neither change in the carrier dynamics indicates that the initial fast process  may not be attributed to electron trapping. With antiphase boundaries present (Figure \ref{fig:fig2}), the origin of the observed initial fast process in LT-GaAs/Si is likely caused by electron scattering (detailed discussion in Section 3.5). And photocarriers can quickly return to the valence band through interaction with the lattice.

\subsection{Supplementary characterization for defect profiling}
\begin{figure}
	\centering
	\includegraphics[width=\linewidth]{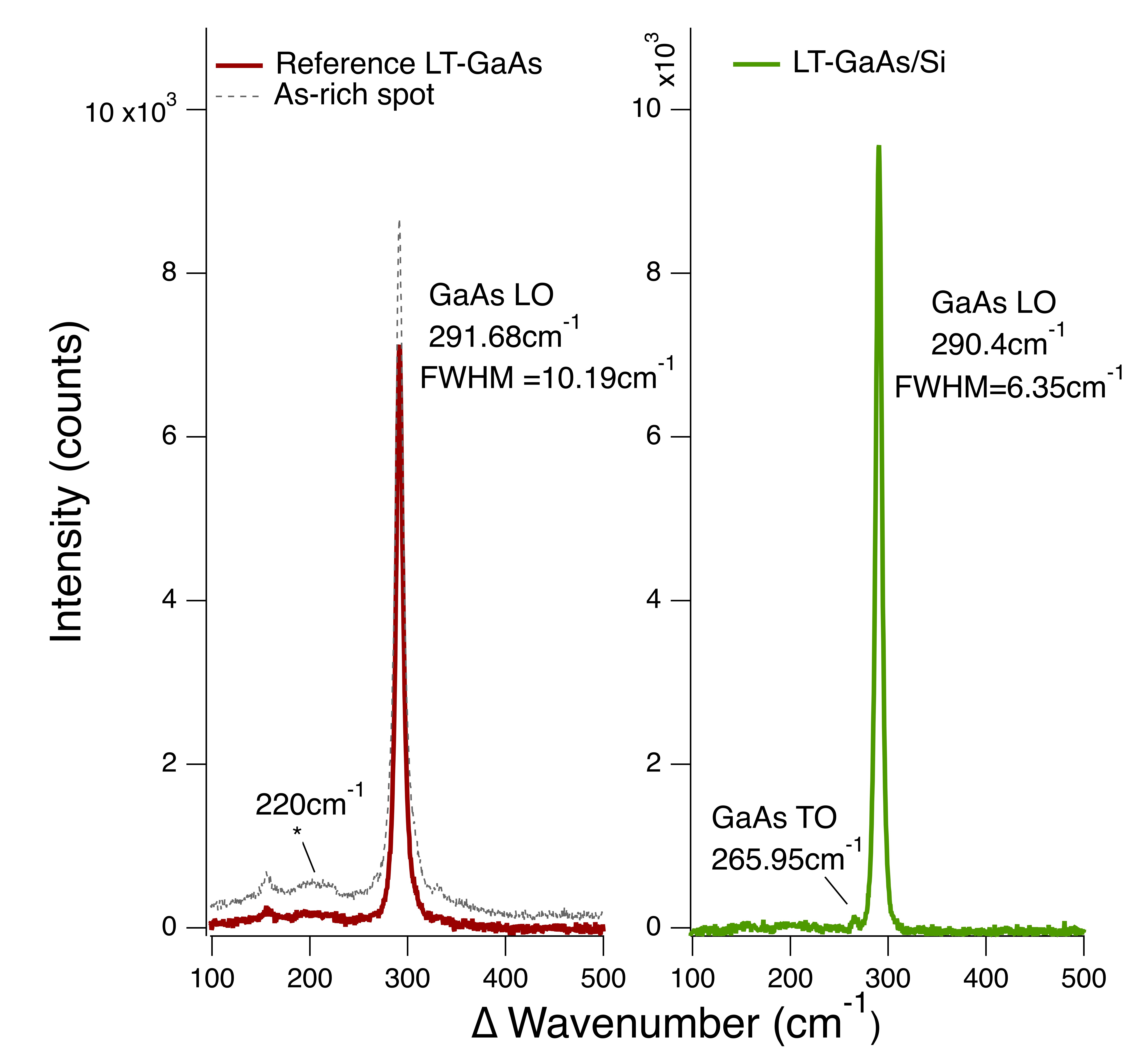}
	\caption{Raman spectra showing GaAs LO phonon peaks for both samples. LT-GaAs/Si shows a small GaAs TO phonon signal. When probed in different spots, the reference sample shows a broad feature around 220 cm$^{-1}$, which may be attributed to As \cite{Abe1996},\cite{Toufella1999}.}
	\label{fig:fig6}
\end{figure}

The Raman spectra obtained from the samples are shown in Figure \ref{fig:fig6}, The GaAs LO peak can clearly be observed (Reference: 291.68 cm$^{-1}$, LT-GaAs/Si: 290.4 cm$^{-1}$), while the GaAs TO peak (265.95 cm$^{-1}$) only appeared for the LT-GaAs/Si sample. Under the selection rules for backscattering geometry, the presence of the Raman-forbidden GaAs TO peak usually signifies some misorientation of the crystal away from $<100>$ \cite{Biellman1983}. The TO-LO intensity ratio signified that the degree of misorientation is low \cite{Abe1996}. Additionally, the GaAs LO phonon peak for the reference exhibited broader FWHM and lower intensity; both are attributed to the successful formation of As-related defects within the LT-GaAs bulk. Raman scans on different spots in the LT-GaAs on GaAs revealed some broad feature around 220 cm$^{-1}$, this has previously been attributed to As \cite{Abe1996},\cite{Toufella1999}. This broad feature was not observed for the LT-GaAs/Si when the Raman measurements were performed. These results are consistent with the behavior of the carrier dynamics. 

\subsection{Heteroepitaxial growth of LT-GaAs/Si}
The surface of nominal Si (100) is comprised of monatomic-sized steps. At the initial stages of epitaxy, As-Si bonds form on the substrate surface, forming an As prelayer \cite{Christou1985}. When As and Ga impinge on these steps as deposition proceeds, the formation of antiphase boundaries is highly probable as As-As or Ga-Ga tend to bond \cite{Houdre1990}. For homoepitaxial growth of GaAs on GaAs (100), at low substrate temperatures, and with As overpressure, more As atoms are readily available to bond with Ga atoms impinging on the surface. As such, Ga-Ga bonding is less likely to occur – but an As-rich surface is established. Therefore, despite similar As overpressure environments, the excess As atoms are incorporated in the crystal differently in each case. Instead of forming point defects with a similar density to that of conventional LT-GaAs, it is possible that most of the excess arsenic atoms participated in the creation of the initial As-Si interface, as well as the succeeding formation of APBs.

It should be emphasized that LT-GaAs/Si still exhibited the favored two-step process that is similar to conventional LT-GaAs. The lack of trap-saturation behavior, however, indicates that the initial fast process may not be attributed to electron trapping alone \cite{Frankel1993}, if at all. The presence of APBs (Figure \ref{fig:fig2}) in the layer must be considered and the fast process could be better attributed to the scattering of conduction electrons on these structural defects.  Minority carrier lifetime decreases when dislocation density increases in GaAs on silicon grown at higher temperatures \cite{Ahrenkiel1990}, \cite{Afalla2019}, and the mechanism is likely to be similar. In THz applications, the lack of trap-saturation behavior may also be favorable, as it implies consistent bandwidths (and possibly dynamic range) when operating LT-GaAs/Si devices regardless of incident laser power. Results of previous work on different LT-GaAs/Si layers showed that whether the LT-GaAs bulk layer is single crystal (i.e. continuous bulk, with As point defects or clusters likely to have formed) or if grain boundaries are present, the desired initial fast decay was still present \cite{Afalla2020},\cite{Afalla2019}. Also, the time constants for these transitions can be well within the desirable range of a few picoseconds. Care has to be taken, as the presence of too many grain boundaries, dislocations, and the like, can impede electron mobility \cite{Afalla2019}, \cite{Wood1982}. For THz detection, the dynamic range increases when the carrier lifetime is faster (regardless of mechanism), but the device is more responsive when the crystal is continuous since the collection of current is much more efficient \cite{Afalla2020},\cite{Frankel1993}. 

\section{Conclusions}
The ultrafast optical properties of heteroepitaxial grown LT-GaAs/Si were investigated and compared with a reference LT-GaAs grown on a SI-GaAs substrate. From coherent phonon spectroscopy, the intrinsic carrier densities were found to be in the order of $~10^{15}$ cm$^{-3}$; under similar excitation densities, we observe a slightly higher total carrier density for the sample grown on Si. Under high fluence values, carrier dynamics measurements showed a trap-saturation behavior in the reference that was not present in the LT-GaAs/Si sample. From SEM images and Raman scattering, the presence of ABPs and slight crystal misorientation were detected for the LT-GaAs/Si sample, while As-rich spots were found for the reference. The difference in defect profiles were attributed to how excess As atoms participated during growth. Although low temperature growth is designed to encourage the excess As to form point defects in the bulk, which manifest electronically as carrier traps, in heteroepitaxy, excess As atoms likely participated in forming APBs instead. The presence of such boundaries can lead to increased carrier scattering. As evidenced by the difference in defect profiles, the initial fast process in the LT-GaAs/Si was attributed instead to carrier scattering.

\ack
A. Salvador acknowledges the Office of the Chancellor of the University of the Philippines Diliman, through the Office of the Vice Chancellor for Research and Development, for funding support through the Outright Research Grant.

\section*{References}
\bibliography{iopart-num}

\providecommand{\newblock}{}
\begin{thebibliography}{10}
\expandafter\ifx\csname url\endcsname\relax
  \def\url#1{{\tt #1}}\fi
\expandafter\ifx\csname urlprefix\endcsname\relax\def\urlprefix{URL }\fi
\providecommand{\eprint}[2][]{\url{#2}}

\bibitem{Petrushkov2020}
Petruskhov M~O, Abramkin D~S, Emelyanov E~A, Putyato M~A, Vasev A~V, Loshkarev
  D~I, Yesin M~Y, Komkov O~S, Firsov D~D and Preobrashenskii V~V 2020 {\em
  Semicond.\/} {\bf 54} 1548

\bibitem{Barrett2019}
Barrett C, Atassi A, Kennon E~L, Weinrich Z, Haynes K, Bao X~Y, Martin P and
  Jones K 2019 {\em J. Mater. Sci.\/} {\bf 54} 1028

\bibitem{Bugomilowicz2016}
Bogumilowicz Y, Harmann J~M, Rochat N, Salaun A, Martin M, Bassani F, Baron T,
  David S, Bao X~Y and Sanchez E 2016 {\em J. Cryst. Growth\/} {\bf 453} 180

\bibitem{Wirths2018}
Wirths Y~S, Mayer B~K, Schmid H, Sousa M, Gooth J, Riel H and Moselund K~E 2012
  {\em ACS Nano\/} {\bf 12} 2169

\bibitem{Bolkhovityanov2008}
Bolkhovityanov Y~B and Pchelyakov O~P 2008 {\em Phys. Uspekhi\/} {\bf 51} 437

\bibitem{Gonzalez1992}
Gonzalez Y, Gonzalez L and Briones F 1992 {\em Jpn. J. Appl. Phys.\/} {\bf 31}
  L816

\bibitem{CGonzales2021}
Gonzales K, Prieto E~A, Catindig G, Reyes A~D~L, Faustino M~A,
  Tumanguil-Quitoras M, Husay H~A, Vasquez J~D, Somintac A, Estacio E and
  Salvador A 2021 {\em J. Mater. Sci. Mater. Electron.\/}  1--12

\bibitem{Afalla2020}
Afalla J, Catindig G, Reyes A~D~L, Prieto E, Faustino M~A, Vistro V~D, Gonzales
  K~C, Bardolaza H, Mag-usara V~K, Husay H~A, Muldera J, Cabello N~I, Ferrolino
  J~P, Kitahara H, Somintac A, Salvador A, Tani M and Estacio E 2020 {\em J.
  Phys. D: Appl. Phys.\/} {\bf 53} 095105

\bibitem{Kamo2014}
Kamo Y, Kitazawa S, Ohshima S and Hosoda Y 2014 {\em Jpn. J. Appl. Phys.\/}
  {\bf 53} 032201

\bibitem{Klier2015}
Klier J, Torosyan G, Schreiner N~S, Molter D, Ellrich F, Zouaghi W, Peytavit E,
  Lampin J~F, Beigang R, Jonuscheit J and von Freymann G 2015 {\em In. J.
  Antenna. Propag.\/} {\bf 2015} 540175

\bibitem{Kasai2009}
Kasai S, Katagiri T, Takayanagi J, Kawase K and Ouchi T 2009 {\em Appl. Phys.
  Lett.\/} {\bf 94} 113505

\bibitem{Youn2008}
Youn D~H, Lee S~H, Ryu H~C, Jung S~Y, Kang S~B, Kwak M~H, Kim S, Choi S~K, Baek
  M~C, Kang K~Y, Kim C~S, Yee K~J, Ji Y~B, Lee E~S, Jeon T~I, Kim S~J, Kumar S
  and Kim G~H 2008 {\em J. Appl. Phys.\/} {\bf 103} 123528

\bibitem{Prahbu1997}
Prahbu S~S, Ralph S~E, Melloch M~R and Harmon E~S 1997 {\em Appl. Phys.
  Lett.\/} {\bf 70} 2421

\bibitem{Dekorsy1993}
Dekorsy T, Kurz H, Zhou X~Q and Ploog K 1993 {\em Appl. Phys. Lett.\/} {\bf 63}
  2899

\bibitem{Georgakilas1992}
Georgakilas A, Panayotatos P, Stoemenos J, Mourrain J~L and Christou A 1992
  {\em J. Appl. Phys.\/} {\bf 71} 2679

\bibitem{Georgakilas1993}
Georgakilas A, Stoemenos J, Tsagaraki T, Komninou P, Flevaris N, Panayotatos P
  and Christou A 1993 {\em J. Mater. Res.\/} {\bf 8} 1908

\bibitem{Alcotte2016}
Alcotte R, Martin M, Moeyaert J, Cipro R, David S, Bassani F, Ducroquet F,
  Bugomilowicz Y, Sanchez E, Ye Z, Bao X~Y, Pin J~B and Baron T 2016 {\em APL
  Mater.\/} {\bf 4} 046101

\bibitem{Koch1987}
Koch S~M, Rosner S~J, Hull R, Yoffe G~W and Harris J~S 1987 {\em J. Cryst.
  Growth\/} {\bf 81} 205

\bibitem{Mooradian1967}
Mooradian A and McWhorter A~L 1967 {\em Phys. Rev. Lett.\/} {\bf 19} 849

\bibitem{Hase1998}
Hase M, Mizoguchi K, Harima H, Miyamaru F, Nakashima S, Fukusawa R, Tani M and
  Sakai K 1998 {\em J. Lumin.\/} {\bf 76,77} 68

\bibitem{Fukusawa1994}
Fukusawa R and Perkowitz S 1994 {\em Phys. Rev. B\/} {\bf 50} 14119

\bibitem{Dekorsy2000}
T~Dekorsy G~C~C and Kurz H 2000 Coherent phonons in condensed media {\em Topics
  in Applied Physics\/} ({\em Light Scattering in Solids VIII\/} vol~76) ed
  Cardona M and Güntherodt G (Springer)

\bibitem{Balgos2019}
Balgos M~H, Jaculbia R, Prieto E~A, Tani M, Estacio E, Salvador A, Somintac A,
  Hayazawa N and Kim Y 2019 {\em J. Appl. Phys.\/} {\bf 126} 235706

\bibitem{Prieto2014}
Prieto E~A~P, Vizcara S~A~B, Somintac A~S, Salvador A~A, Estacio E~S, Que C~T,
  Yamamoto K and Tani M 2014 {\em J. Opt. Soc. Am. B\/} {\bf 31} 291

\bibitem{Segschneider2002}
Segschneider G, Jacob F, Loffler T, Roskos H~G, Tautz S, Kiesel P and Dohler G
  2002 {\em Phys. Rev. B\/} {\bf 65} 125205

\bibitem{Nemec2001}
Nemec A, Pashkin A, Kužel P, Khazan M, Schnüll S and Wilke I 2001 {\em J.
  Appl. Phys.\/} {\bf 90} 1303

\bibitem{Sosnowski1997}
Sosnowski T~S, Norris T~B, Wang H~H, Grenier P, Whitaker J~F and Sung C~Y 1997
  {\em Appl. Phys. Lett.\/} {\bf 70} 3245

\bibitem{Savard2010}
Savard S, FAllard J, Bernier M, Petersen J~C, Dodge J~S, Fournier P and Morris
  D 2010 {\em J. Appl. Phys.\/} {\bf 108} 124507

\bibitem{Biellman1983}
Biellman J, Prevot B and Schwab C 1983 {\em J. Phys. C: Solid State Phys.\/}
  {\bf 16} 1135

\bibitem{Abe1996}
Abe H, Harima H, Nakashima S, M~Tani K~S, Tokuda Y, Kanamoto K and Abe Y 1996
  {\em Jpn. J. Appl. Phys.\/} {\bf 35} 5955

\bibitem{Toufella1999}
Toufella M, Puechh P, Carles R, Bedel E, Fontaine C, Claverie A and Benassayag
  G 1999 {\em J. Appl. Phys.\/} {\bf 85} 2929

\bibitem{Christou1985}
Christou A, Wilkins B~R and Tseng W~F 1985 {\em Electron. Lett.\/} {\bf 21} 406

\bibitem{Houdre1990}
Houdré H and Morkoç H 1990 {\em Crit. Rev. Solid State Mater. Sci.\/} {\bf
  16} 92

\bibitem{Frankel1993}
Frankel M, Tadayon B and Carruthers T~F 1993 {\em Appl. Phys. Lett.\/} {\bf 62}
  255

\bibitem{Ahrenkiel1990}
Ahrenkiel R~K, Al-Jassim M~M, B~Keyes D~D, Jones K~M, Vernon S~M and Dixon T~M
  1990 {\em J. Electrochem. Soc.\/} {\bf 137} 996

\bibitem{Afalla2019}
Afalla J, Gonzales K~C, Prieto E~A, Catindig G, Vasquez J~D, Husay H~A,
  Tumanguil-Quitoras M~A, Muldera J, Kitahara H, Somintac A, Salvador A,
  Estacio E and Tani M 2019 {\em Semicond. Sci. Technol.\/} {\bf 34} 035031

\bibitem{Wood1982}
Wood J, Howes M~J and Morgan D~V 1982 {\em Phys. Stat. Sol. A\/} {\bf 74} 493

\end{thebibliography}

\end{document}